\documentclass[%
reprint,
superscriptaddress,
amsmath,amssymb,
aps,
prl,
floatfix
]{revtex4-2}

\usepackage{multirow}
\usepackage{graphicx}
\usepackage{dcolumn}
\usepackage{bm}
\usepackage[version=3]{mhchem}
\usepackage{nicefrac}

\usepackage{color}
\usepackage{xcolor}
\usepackage{dcolumn}
\usepackage{amssymb}
\usepackage{url}
\usepackage{hyperref}
\usepackage{xfrac}
\usepackage{tabularx}
\usepackage{xspace}
\usepackage{amsmath}
\usepackage[normalem]{ulem}
\usepackage{mathtools}
\usepackage{physics}
\usepackage{dsfont}
\usepackage{float}
\usepackage{braket}
\usepackage{placeins}

\def\be{\begin{equation}}
\def\ee{\end{equation}}
\def\bea{\begin{eqnarray}}
\def\eea{\end{eqnarray}}
\def\ba{\begin{array}}
\def\ea{\end{array}}

\newcommand{\Alm}{A\textit{l}M}

\begin{document}

\title{Zoology of Altermagnetic-type Non-collinear Magnets on the Maple Leaf Lattice}
\author{Pratyay Ghosh}
\email{pratyay.ghosh@epfl.ch}
\affiliation{Institute of Physics, Ecole Polytechnique Fédérale de Lausanne (EPFL), CH-1015 Lausanne, Switzerland}
\author{Ronny Thomale}
\email{ronny.thomale@uni-wuerzburg.de}
\affiliation{Institut f\"ur Theoretische Physik und Astrophysik and W\"urzburg-Dresden Cluster of Excellence ct.qmat, Universit\"at W\"urzburg, Am Hubland Campus S\"ud, W\"urzburg 97074, Germany}

\begin{abstract}
We define unconventional non-collinear magnetic ground states on the maple leaf lattice (MLL) distinguished by the selective breaking or preservation of time reversal ($\mathcal{T}$) and parity ($\mathcal{P}$). Depending on the nature of $\mathcal{P}\mathcal{T}$-breaking, linear spin-wave theory reveals momentum-dependent non-relativistic magnon spin splitting at different high symmetry points in the Brillouin zone. From a mean-field analysis of the Hubbard model at weak coupling, we reveal itinerant $\mathcal{P}$-preserving $q=0$ altermagnetic (\Alm)-type order, while we expect $\mathcal{P}$-broken canted-$120^\circ$ \Alm-type order at strong coupling. Our findings establish the MLL as a prime platform for exploring phase transitions and frustration phenomena emanating from competing non-collinear \Alm-type orders.
\end{abstract}

\maketitle

\textit{Introduction.}-- Altermagnetism (\Alm)~\cite{PhysRevX.12.031042,Smejkal2022,PhysRevX.12.040002} has recently been introduced as a class of collinear magnets with crystal-symmetry compensated magnetization and non-relativistically spin-split~\cite{mejkal2020,Naka2019,PhysRevLett.131.256703} magnon bands. Given that such spin splitting would not be limited to relativistic corrections such as spin-orbit coupling, the \Alm's promise is to provide a favorably operable spintronics material platform where spin-polarized magnons can be utilized for magnetic signal processing and transmission. Even more drastically than the prospect of its potential technological use, however, \Alm~has revitalized the research domain of quantum antiferromagnetism to carefully revisit the nature of itinerant and localized magnets from the viewpoint of crystal symmetry and associated degrees of freedom such as orbitals~\cite{PhysRevLett.132.236701} and sublattices~\cite{2g3v-z76q}, as well as symmetries of non-crystallographic nature~\cite{durrnagel2025extendedswavealtermagnets}. Classifying \Alm~order from a spin group theoretical perspective proved particularly feasible due to their collinear magnetic nature, where the deconstructionist assembly principle of \Alm s amounts to connecting staggered magnetic moments through crystal symmetry operations, but not translation symmetry in order to evade any Kramer's band degeneracy. As one includes non-collinearity of spins, however, the classification of \Alm-type magnets, or rather unconventional magnets in general, necessitates a more principled approach~\cite{classification1,classification2,classification3} still to be fully accomplished, and enforces to deviate from the precise classification nomenclature of hitherto defined \Alm s~\cite{jungwirth2025symmetrymicroscopyspectroscopysignatures,sym_class,hellenes2024pwavemagnets}. While it falls short of an exhaustive classification since it is not a sufficient condition to reveal \Alm-type signatures in an unconventional non-collinear magnet~\cite{PhysRevB.109.024404}, a still useful defining subcategorizing property can be associated with the nature of $\mathcal{P}\mathcal{T}$ breaking~\cite{Cheong2024,Cheong2025,JUNGWIRTH2025100162}, where $\mathcal{P}$ and $\mathcal{T}$ denote parity and time-reversal symmetry, respectively: (i) broken $\mathcal{P}$ and broken $\mathcal{T}$, (ii) unbroken $\mathcal{P}$ and broken $\mathcal{T}$, and (iii) broken $\mathcal{P}$ and unbroken $\mathcal{T}$. Similar to \Alm s, non-collinear unconventional magnets can exhibit a variety of exotic spin-dependent phenomena such as higher-order anomalous Hall effect (AHE)~\cite{PhysRevB.109.104413,Kbler2014,Nakatsuji2015,oddparity_exp}, piezomagnetism~\cite{PhysRevB.110.214428,PhysRevMaterials.8.L041402,Ikhlas2022}, and current-induced magnetization the further characterization of which depends on the specific breaking of $\mathcal{P}$ and/or $\mathcal{T}$. 

 It is rather suggestive that lattices with a sufficient amount of sublattices and/or orbitals are necessary to implement staggered magnetization patterns connected through spin group symmetry generators, which can likewise lead up to the sublattice/orbital-decoration of other electronic orders such as superconductivity~\cite{PhysRevB.110.024501}. While a systematic construction principle of \Alm-type non-collinear magnets  is less apparent, the rudimentary notion of non-relativistic spin splitting in non-collinear antiferromagnets even predates the inception of \Alm~\cite{PhysRevLett.119.187204,PhysRevB.101.220403}. Similarly, at a time when the field was still in its early-born state, we had found~\cite{Ghosh2022} that the Heisenberg model on the maple leaf lattice (MLL) harbors an intriguingly canted, hence non-collinear $120^\circ$ magnet, which by present day nomenclature could be labelled as non-collinear \Alm-type magnet.
The MLL is an Archimedean lattice first introduced in the literature as a depleted triangular lattice in 1995~\cite{Betts1995,Schulenburg2000,Schmalfuss2002,Farnell2011,Farnell2014}. Our earlier work~\cite{Ghosh2022} identified it as the only other lattice model with uniform tilings besides the Shastry–Sutherland model from 1981~\cite{Shastry1981} to host an exact dimer product ground state, which has triggered recent research activity~\cite{Ghosh2023,Gresista2023,Sonnenschein_QSL_MLM_2024,Beck2024,Ghosh2024-al,PhysRevB.110.085151,Ghosh2025,Penc_MLL,nyckees2025tensornetworkstudygroundstate,Ghosh2023b,Lake2025,Ghosh2025,Schmoll2025,Hutak2025,Ryd_MLL,ebert2026competingparamagneticphasesmapleleaf}. 

\begin{figure*}
    \centering
    \includegraphics[width=0.95\linewidth]{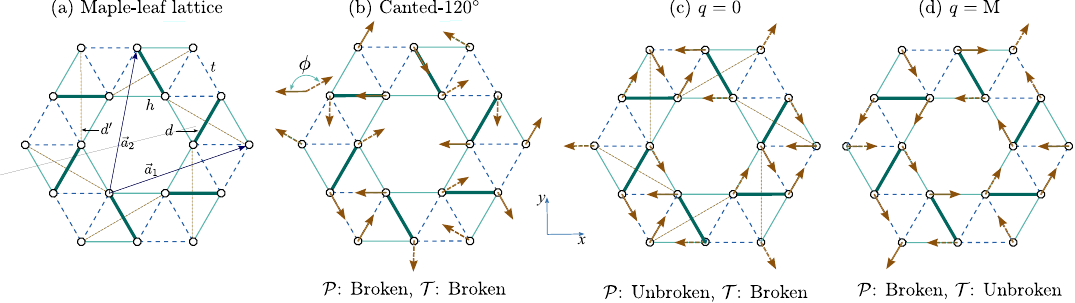}
    \caption{(a) The maple leaf lattice (MLL) exhibits three types of nearest neighbor bonds, $h$, $t$, and $d$, and a subset of second neighbor bonds, $d'$. The bonds correspond to exchange couplings $J_{r}$ or tight-binding hopping amplitudes $t_{r}$, with $r \in \{h,t,d,d'\}$ (lattice vectors indicated). 
    (b)-(d) Three distinct non-collinear, coplanar \Alm~orders: (b) Canted-$120^\circ$; (c) $q=0$; 
    (d) $q = M$.
    }
    \label{fig-orders}
\end{figure*}

In this Letter, we show how the MLL (Fig.~\ref{fig-orders}) provides an ideal platform for realizing non-collinear unconventional magnets with \Alm-type signatures, and quantum frustration phenomena associated with them. We formulate \Alm-type states for all non-collinear subcategories from $\mathcal{P}\mathcal{T}$ breaking. Particularizing to a type (i) state with broken $\mathcal{P}$ and a type (ii) state with preserved $\mathcal{P}$, we analyze their magnon bands and momentum-dependent spin orientation profile. While the type (i) state is stabilized in the localized strong coupling limit of the Hubbard model on the MLL, we find the type (ii) state to be favored at Hartree-Fock level for the itinerant weak coupling limit. Our observations suggest the MLL to be an arena for investigating aspects of criticality, frustration, and exotic descendant order from a parent scenario of competing non-collinear \Alm-type orders.

\textit{Non-collinear unconventional magnets with \Alm~signatures.}--
The MLL is generated from the snub trihexagonal tiling by placing a site at each vertex; the resulting sites occupy the general Wyckoff position $6d$ of the wallpaper group $p6$, characterized by sixfold rotational symmetry and the absence of mirror and glide reflections.
Owing to this symmetry structure, the nearest-neighbor (NN) bonds of the lattice are not all symmetry-equivalent. As it turns out, the MLL is composed of three distinct elements formed by the NN bonds [Fig.~\ref{fig-orders}(a)]: hexamers, $h$ (thin solid lines), trimers, $t$ (dashed lines), and dimers, $d$ (thick solid lines). Apart from the sixfold rotation through the center of these hexamers, the MLL also possesses a threefold rotation center at the center of each trimer and a twofold rotation center at the midpoint of each dimer, but no mirror symmetry. The action of space group operations decouples into subblocks of the same structural motif.
The limited symmetry equivalence in MLL enlarges the number of independent magnetic degrees of freedom and allows several symmetry-allowed magnetic orders to be realized on the lattice. Let us assume coplanar magnetic order and a momentum-independent spin quantization axis in the following. For our purpose, it is sufficient to discuss only three cases which break $\mathcal{PT}$, or, more precisely, $\mathcal{PT}\tau$, where $\tau$ denotes translation, which is a necessary condition for any \Alm-type signatures. Why that is becomes apparent from the spin-resolved dispersion $\varepsilon(\mathbf{k},\mathbf{s})$ and how it transforms under $\mathcal{P}$, $\mathcal{T}$, and $\tau$: $\mathcal{T}\varepsilon(\mathbf{k},\mathbf{s})\mathcal{T}^{-1}=\varepsilon(-\mathbf{k},-\mathbf{s})$; $\mathcal{P}\varepsilon(\mathbf{k},\mathbf{s})\mathcal{P}^{-1}=\varepsilon(-\mathbf{k},\mathbf{s})$; $\tau\varepsilon(\mathbf{k},\mathbf{s})\tau^{-1}=\varepsilon(\mathbf{k},\mathbf{s})$. Since a conserved $\mathcal{PT}\tau$ enforces spin-degenerate bands at all $\mathbf{k}$, breaking $\mathcal{PT}\tau$ is necessary for spin splitting. Furthermore, the magnetic order must also lack $U(\pi)\tau$ symmetry, where $U(\theta)$ is a spin-$1/2$ spinor rotation by angle $\theta$ around the axis perpendicular to the spin plane  $U(\pi)\varepsilon(\mathbf{k},\mathbf{s})U(\pi)^{-1}=\varepsilon(\mathbf{k},\mathbf{-s})$, so that reversing all spins shall not be restorable by translation~\cite{PhysRevMaterials.5.014409,Smejkal2022}. 

We now identify non-collinear magnetic orders on the MLL falling into the type (i)-(iii) subcategories. The first is the classical magnetic order known as the canted-$120^\circ$ order emerging from a Heisenberg model on the MLL~\cite{Farnell2011,Ghosh2022}. In this configuration, each trimer adopts a local $120^\circ$ spin arrangement, while a non-local spin canting exists between the neighboring trimers. The canting angle, $\phi$, depends on the competition between the interactions associated with the hexamer and the dimer bonds \cite{SM}. An example of this magnetic order is shown in Fig.~\ref{fig-orders}(b). This order breaks translation symmetry, reduces the $C_{6z}$ symmetry about the hexamer centers to $C_{3z}$ and breaks both $\mathcal{P}$ and $\mathcal{T}$ symmetries, thus rendering it a natural realization of a type (i) state. We establish a type (ii) state through the magnetic pattern shown in Fig.~\ref{fig-orders}(c). Unlike the canted-$120^\circ$ order, this configuration preserves $\mathcal{P}$, which derives from the $C_{2z}$ symmetry about the hexamer centers, as well as through the midpoint of the dimers. Additionally, the spin configuration exhibits a combined $U(\pi)R_{6z}$ symmetry, where 
$R_{6z}$ denotes a sixfold spatial rotation combined with a spin rotation. We henceforth refer to this configuration as the `$q=0$ order', reflecting the preservation of translational symmetry. The type (iii) state we propose is characterized by an ordering wavevector $q=M$ [Fig.~\ref{fig-orders}(d)]. This state breaks $\mathcal{P}$ and translation symmetry $\tau$ along $\vec{a}_1$, while it preserves $\mathcal{T}\tau$. 

Due to their distinct symmetry properties, these non-collinear magnets can be expected to show different physical responses. 
The $q=0$ order preserves $\mathcal{P}$, resulting in an even-parity spin splitting. This allows for transverse, even-order current-induced magnetization, with the applied current and the induced magnetization both lying in the spin plane, i.e., high-odd-order AHE. 
By contrast, the canted-$120^\circ$ order, as the $q=\mathrm{M}$ order, breaks inversion symmetry $\mathcal{P}$, which allows $\varepsilon(\mathbf{k},\mathbf{s}) \ne \varepsilon(-\mathbf{k},\mathbf{s})$. 
Since the $q=\mathrm{M}$ order preserves time-reversal symmetry $\mathcal{T}$ as well as the combined symmetry $U(\pi)\mathcal{T}$,
$\varepsilon(\mathbf{k},\mathbf{s}^\perp)=\varepsilon(-\mathbf{k},-\mathbf{s}^\perp)$; that is, the coplanar order generates an odd-parity collinear spin polarization $\mathbf{s}^\perp$ in momentum space orthogonal to the spin plane of the magnetic order, and zero inplane response~\cite{hellenes2024pwavemagnets,sym_class}.
Consequently, our $q=\mathrm{M}$ state can exhibit an even-order AHE for an applied current perpendicular to the spin plane. 
The canted-$120^\circ$ order, in contrast, does not preserve $\mathcal{T}$ and therefore allows $\varepsilon(\mathbf{k},\mathbf{s}) \neq \varepsilon(-\mathbf{k},-\mathbf{s})$. However, it does preserve $U(\pi)\mathcal{T}$, which enforces
$\varepsilon(\mathbf{k},\mathbf{s}^\parallel)=\varepsilon(-\mathbf{k},\mathbf{s}^\parallel)$,
where $\mathbf{s}^\parallel$ denotes the in-plane spin polarization. Any in-plane spin splitting must thus be of even parity~\cite{Zhang2018}, whereas no constraint is implied for  $\mathbf{s}^\perp$ leading to the possibility of mixed even and odd parity responses for the out-of-plane component. Thus, the type (i) canted-$120^\circ$ state can exhibit mixed characteristics of type (ii) and type (iii).  

\textit{Microscopic realization at strong coupling.}--
Owing to the symmetry inequivalence of the three types of bonds in the MLL, one can assign distinct couplings to each bond type and define a Heisenberg Hamiltonian as
\begin{math}
    H=\sum_{\langle ij\rangle_r} J_r \mathbf{S}_i\cdot\mathbf{S}_j,
\end{math}
where $\langle ij\rangle_r$ sums over neighbors connected by a bond of type $r$, with $r \in \{h,t,d\}$, and $J_r$ denoting the corresponding coupling strength [Fig.~\ref{fig-orders}(a)]. The canted-$120^\circ$ configuration was initially identified as the classical ($S\to\infty$) ground state for the purely antiferromagnetic case with $J_h = J_t$~\cite{Farnell2011,Ghosh2022}. This magnetic structure, however, remains stable across a wide region of parameter space even when $J_h \ne J_t$~\cite{Ghosh2025,gresista2025} and down to the quantum limit $S=1/2$, where the interplay of geometric frustration and quantum fluctuations favors the canted-$120^\circ$ order for $J_h = J_t \gtrsim 0.7 J_d$~\cite{Farnell2011,Ghosh2022,Beck2024,Ghosh2023b,nyckees2025tensornetworkstudygroundstate}.
The realization of the $q=0$ order requires the inclusion of a subset of second-neighbor bonds, denoted $d'$ in Fig.~\ref{fig-orders}(a)~\cite{Ghosh2025}, to which we assign a Heisenberg coupling of strength $J_{d'}$. 
For classical vector spins, the $q=0$ order can be stabilized by choosing antiferromagnetic $J_t$ and $J_h$ together with ferromagnetic $J_d$ and $J_{d'}$, subject to the condition
\begin{math}
   J_t \ge J_h, \quad |J_{d'}| > J_h^2/|J_d|
\end{math}~\cite{SM}. 
The $q=M$ order has not been previously reported.
Here, individual hexamers can adopt two distinct spin configurations. Along a strip in the $\vec{a}_2$ direction, each hexamer assumes the same configuration, while the configuration alternates along the $\vec{a}_1$ direction. Treating each hexamer as an effective supersite and the two configurations as a two-level system, one sees that the $q=M$ order effectively maps onto a stripe magnetic order on the triangular lattice, which is known to be stabilized in the antiferromagnetic $J_1$-$J_2$ Heisenberg model for sufficiently strong $J_2$~\cite{tri_1,tri_2,tri_3}. By analogy, we speculate that the stabilization of the $q=M$ order in the MLL requires further-neighbor inter-hexamer interactions, which we defer to a later point.

\begin{figure}
    \centering
    \includegraphics[width=0.95\linewidth]{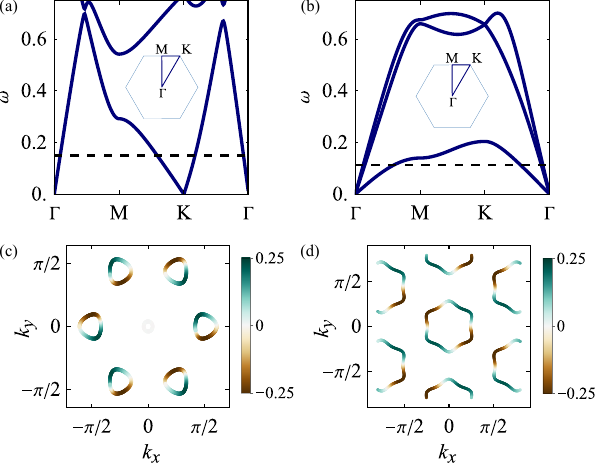}
    \caption{Spin-wave spectrum of (a) canted-$120^\circ$ order for $(J_h,J_t,J_d)=(1.0,1.0,1.5)$ and  (b) $q=0$ order for $(J_h,J_t,J_d,J_{d'})=(0.4,0.6,-1.0,-0.25)$. (c), (d) $\langle S^\parallel\rangle$ along constant-energy contours [dashed line in panels (a) and (b)] for the lowest band of the canted-$120^\circ$ (c),(d) Spin splitting around $K, K'$ in (a) and $\Gamma$ in (b). Orders (a) and (b) show parity-even spin momentum singatures.}
    \label{fig-LSWT}
\end{figure}

\textit{Comparison of type (i) and (ii) magnon spectra.}--
Using linear spin-wave theory (LSWT), we observe momentum-dependent spin splitting in the magnon spectrum for the type (i) canted-$120^\circ$ and the type (ii) $q=0$ state. The spin quantization axes are chosen to align with the local orientations of the spins $\mathbf{S}^\parallel_i$ for the classical magnetic order. In this local frame, the usual Holstein-Primakoff transformation maps the spin Hamiltonian onto a quadratic bosonic form. Diagonalizing this Hamiltonian in momentum space via a Bogoliubov transformation yields six magnon branches.
Examples of the low-energy sector of the spin-wave spectrum for both magnetic orders are shown in Figs.~\ref{fig-LSWT}(a),(b). For the canted-$120^\circ$ order, we present results for $(J_h,J_t,J_d) = (1.0, 1.0, 1.5)$, while for the $q=0$ order we take $(J_h,J_t,J_d,J_{d'}) = (0.4, 0.6, -1.0, -0.25)$. The excitation spectra of the magnon gases exhibit three zero modes each, as required by exhaustive SU(2) spin-rotation symmetry breaking. In the canted-$120^\circ$ configuration, we find the acoustic branch reaching zero at the Brillouin zone (BZ) center $\mathbf{k}=\Gamma$ and at the zone corners $\pm\text{K}$ due to the tripling of the magnetic unit cell. The $\mathbf{k}=0$ mode describes out-of-plane spin fluctuations, while $\mathbf{k}=\pm \text{K}$ modes correspond to in-plane oscillations~\cite{Chubukov1994,Schmalfuss2002}. The $q=0$ order exhibits three acoustic and three optical branches, with all acoustic branches evolving from  $\Gamma$. Within LSWT we compute $\langle S^\parallel \rangle$, the spin projection along the local quantization axis, for each magnon band, which we can use to observe the momentum-dependent spin splitting in type (i) and type (ii) magnets. The results for the lowest magnon band in both cases, along constant-energy cuts, are presented in Fig.~\ref{fig-LSWT}(c) and (d). In both cases, the spin splitting exhibits even parity in agreement with the preceding symmetry analysis. For the $q=0$ order, the splitting is most pronounced near the $\Gamma$ point, whereas for the canted-$120^\circ$ order it is strongest near the $\mathrm{K}$ points, corresponding to in-plane fluctuations.

\begin{figure*}
    \centering
    \includegraphics[width=0.95\linewidth]{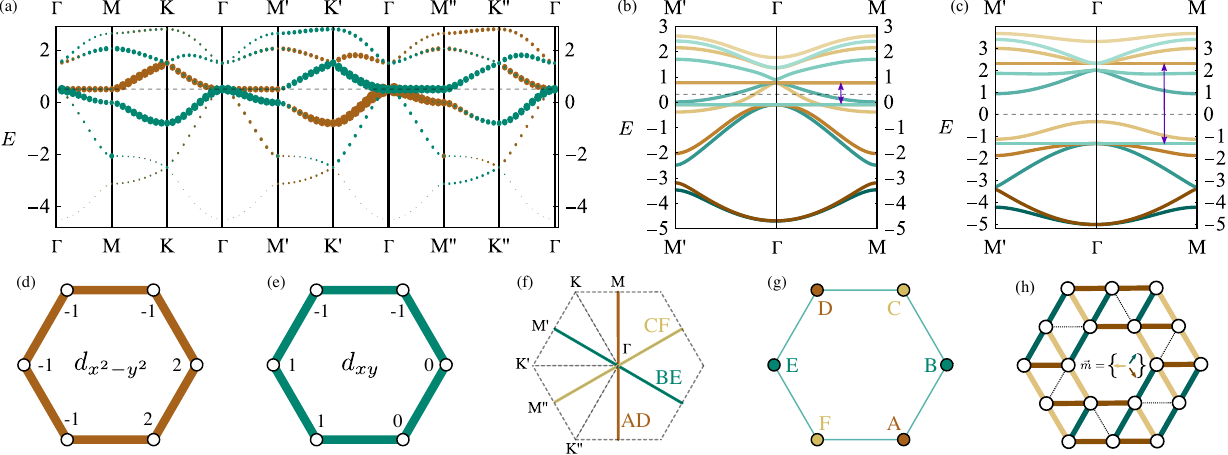}
    \caption{Band structure of \eqref{eq:FH} with $(t_h, t_t, t_d) = (1, 1, 0.5)$ for (a) $U=0$, (b) $U=3$ (altermagnetic metal) and (c) $U=5$ (altermagnetic insulator), plotted along the high-symmetry paths in the BZ.
    The gray dashed lines mark the Fermi level at half filling. (a) The dot sizes indicate the projection of the Bloch states onto the $d$-orbitals shown in panels~(d) and (e) (the numbers specify the relative strength of the sublattice contributions). 
    (b),(c) Double-headed arrows indicate the rigid splitting of the partially flat bands for an altermagnetic metal and insulator.
    (f) Sublattice resolution of the bands crossing the Fermi level for $U=0$ in~\ref{eq:FH}. The corresponding Bloch states are predominantly localized on sublattices at opposite corners of the hexamers, indicated by the same colors in panel~(g). 
    (h) Expectation values  $\langle c_{i,\vec{m}}^\dagger c_{j,\vec{m}} + c_{j,\vec{m}}^\dagger c_{i,\vec{m}} \rangle_0$, on NN bonds for $U=5$. $\vec{m}$ denotes the three constituent magnetic moments, with only the dominant contributions shown. 
    }
    \label{fig-hubbard}
\end{figure*}

\textit{Microscopic realization at weak coupling.}-- We investigate the Hubbard model on the MLL  
\begin{align}\label{eq:FH}
H = - \sum_{\langle ij \rangle_r} \sum_\sigma t_r \left( c_{i\sigma}^\dagger c_{j\sigma} + \text{H.c.} \right)
    + U \sum_i n_{i\uparrow} n_{i\downarrow}, 
\end{align}
where $\langle ij\rangle_r$ sums over the NN bond of type $r$ with hopping amplitude $t_r$. Here, $c_{i\sigma}^\dagger$ ($c_{i\sigma}$) creates (annihilates) an electron with spin $\sigma \in \{\uparrow,\downarrow\}$ at site $i$, and $n_{i\sigma} = c_{i\sigma}^\dagger c_{i\sigma}$ is the density operator. The non-interacting band structure of \eqref{eq:FH} is depicted in Fig.~\ref{fig-hubbard}(a), where the ensuing analysis is inspired by a similar study by Ferrari and Valenti for the Shastry-Sutherland lattice~\cite{PhysRevB.110.205140}.
Noticeably, the fourth band counted from the bottom remains flat along the axial $\Gamma\text{M}$, $\Gamma\text{M}'$, and $\Gamma\text{M}''$ directions in the BZ.
We calculate the projections of the electronic states onto the $s$, $p$, $d$, and $f$ orbital states. The lowest, the second-lowest, and the two uppermost bands are primarily localized on the $s$, $f$, and $p$ orbitals, respectively. The third and the fourth band are found to be mainly composed of $d_{x^2-y^2}$ and $d_{xy}$ orbital contributions, where the $d$-orbital projection of the associated density of states is shown in Fig.~\ref{fig-hubbard}(a) and the dot sizes indicate the relative weights of the Bloch states with respect to the two inplane $d$ orbitals.
The definition of the $d_{x^2-y^2}$ and $d_{xy}$ orbitals can be read off from Fig.~\ref{fig-hubbard}(d) and (e), respectively.
The partially flat bands along the three axial directions of the BZ, which lie at the Fermi level at half-filling, correspond to Bloch states localized at three different sublattice pairs. Each of these pairs is formed by opposite sites of a hexamer and related to the others by threefold rotations, as illustrated in Fig.~\ref{fig-hubbard}(f) and (g).
We treat the Hubbard on-site interaction $U \sum_i n_{i\uparrow} n_{i\downarrow}$, through a spin-rotation–invariant mean-field (Hartree–Fock) decoupling. We define the local magnetization at each site as
$
\mathbf{m}_i = \langle c_{i\alpha}^\dagger \, \boldsymbol{\sigma}_{\alpha\beta} \, c_{i\beta} \rangle
$, allowing to approximate the interaction term by
\begin{equation}\label{eq-mf}
U \, n_{i\uparrow} n_{i\downarrow}
\;\approx\;
-\,\frac{U}{2}\,\mathbf{m}_i \cdot 
\sum_{\alpha,\beta} c_{i\alpha}^\dagger \, \boldsymbol{\sigma}_{\alpha\beta} \, c_{i\beta}
\;+\;
\frac{U}{4}\,|\mathbf{m}_i|^2,
\end{equation}
where $\boldsymbol{\sigma}$ is the vector of Pauli matrices. At zero temperature, the mean-field energy is computed by summing all eigenvalues $\le 0$, and $\mathbf{m}_i$ is determined self-consistently by minimizing the energy in momentum space.

As a progression for increasing $U$ the system undergoes a transition from the paramagnetic metal phase to an \Alm-type metallic phase, and eventually into an \Alm-type insulator phase where throughout the system favours $q=0$ \Alm-type order. Figures~\ref{fig-hubbard}(b) and (c) show the electronic band structure at half-filling for $(t_h, t_t, t_d) = (1, 1, 0.5)$, with $U=3$ (metallic) and $U=5$ (insulating). In both cases, the system exhibits $d$-wave \Alm-type band splitting.
The splitting originates from the localization of the paramagnetic bands on different sublattice pairs, each formed by opposite sites of a hexamer in the noninteracting case. Note that in the $q=0$ order, the opposite sites of a hexamer have spins aligned in the same direction. The first term in the mean-field decoupling of Eq.~\eqref{eq-mf} acts as a site-dependent Zeeman field, with $\mathbf{m}_i$ serving as an effective magnetic field. Bands localized on a given pair of sublattices are thus split according to their corresponding spin orientation. As a result, a pronounced rigid splitting occurs for the third and fourth bands along the axial directions of the BZ, where the nonmagnetic states are predominantly localized on the $d_{x^2-y^2}$ and $d_{xy}$ orbitals.
 To probe a real space signature of this, we compute spin-dependent kinetic terms on the NN bonds, $\langle c_{i,\vec{m}}^\dagger c_{j,\vec{m}}+c_{j,\vec{m}}^\dagger c_{i,\vec{m}}\rangle$ and present a schematic summary in Fig.~\ref{fig-hubbard}(h). There, $\vec{m}$ indicates the three species of spin moments in the $q=0$ order, oriented at $120^\circ$ with respect to each other. We observe distinct behavior of the three sets of rhombic plaquettes containing a dimer bond: each set is characterized by 
strong kinetic terms involving two of the three $\vec{m}$ species, with the opposite rhombuses showing the same pattern. This $d$-wave-like pattern in real space mirrors the $d$-wave spin splitting seen in momentum space.
        
\textit{Discussion.}--
We have shown how the MLL supports various distinct classes of fully compensated non-collinear \Alm-type magnetic orders in the strongly coupled Heisenberg model limit. It naturally provokes the identification and analysis of a frustrated parametric regime with competing \Alm-type magnets as future work, in particular with regard to descending electronic order from a frustrated \Alm~parent regime as well as the formation of \Alm~spin liquids as a manifestation of topological order and fractionalized spinon quasiparticles~\cite{PhysRevResearch.7.023152}. Similarly, a frustrated \Alm-type magnetic regime is likely to unfold at intermediate Hubbard coupling, where the type (ii) $q=0$ state from weak coupling and the type (i) canted-$120^\circ$ state from strong coupling face off, suggesting itself for future investigation. 
At the level of a possible experimental realization of \Alm-type magnets on MLL, \emph{ab initio} and density functional theory studies of maple-leaf compounds are starting to  provide experimentally tangible material-specific predictions~\cite{Guo2025}. The distinguishability of our different types of non-collinear \Alm-type order can be conveniently achieved through the distinct momentum-space locations of spin (band) splitting, accessible through various density of states spectroscopy or, most directly, via inelastic neutron scattering~\cite{PhysRevB.108.L180401}, so is the predicted $d$-wave spin texture via spin-resolved ARPES~\cite{PhysRevLett.132.036702} along with the intrinsic spin Hall effect  through spin Hall transport measurements~\cite{SM,RevModPhys.87.1213}.

\textit{Acknowledgments}--
We thank D. Agterberg, R. Fernandes, J. Sinova, L. \v{S}mejkal, and R. Valenti for discussions. PG acknowledges financial support through the Swiss National Funds. RT acknowledges
financial support by the Deutsche Forschungsgemeinschaft through Project-ID 258499086 – SFB 1170, through the W\"urzburg- Dresden Cluster of Excellence on Complexity and Topology in Quantum Matter – ctd.qmat Project-ID 390858490 – EXC 2147, and through the Research Unit QUAST, Project-ID 449872909 – FOR5249.

\bibliography{Refs}
\end{document}